\documentclass[aps,prl,twocolumn,floatfix, preprintnumbers]{revtex4}
\usepackage[pdftex]{graphicx}
\usepackage[mathscr]{eucal}
\usepackage[pdftex]{hyperref}
\usepackage{amsmath,amssymb,bm}

\newif\ifarxiv
\arxivfalse
\begin		{document}

\def\Nc		{N_{\rm c}}

\def\section	#1{\quad\textit{#1}.---}
\def\suck[#1]#2{\includegraphics[#1]{#2}}        % for arXiv

%\preprint{}

\title
    {
    Colliding shock waves and hydrodynamics in small systems
    }

\author{Paul~M.~Chesler}
\affiliation
    {Department of Physics, Harvard University, Cambridge, MA 02138, USA}
\email{pchesler@physics.harvard.edu}

\date{\today}

\begin{abstract}
    Using numerical holography, we study the collision
    of a planar sheet of energy with a bounded localized distribution of energy.
    The collision, which mimics proton-nucleus collisions, produces a localized lump of debris 
    with transverse size $R \sim 1/T_{\rm eff}$ with $T_{\rm eff}$ the effective temperature, and has large gradients and large transverse flow.
    Nevertheless, the postcollision evolution 
    is well described by viscous hydrodynamics.  Our results bolster the notion that debris produced in proton-nucleus collisions 
    may be modeled using hydrodynamics.
\end{abstract}


\maketitle
\parskip	2pt plus 1pt minus 1pt

\noindent \section{{Introduction}}Data on recent light-heavy ion collisions, including proton-nucleus collisions,
indicate the presence of collective flow \cite{Chatrchyan:2013nka,Abelev:2012ola,Aad:2012gla,Adare:2013piz}.
While alternative mechanisms for the origin of the flow have been proposed \cite{Dumitru:2014yza},
the data are consistent with hydrodynamic evolution \cite{Bozek:2011if,Bozek:2012gr,Bzdak:2013zma}.
However, at experimentally accessible energies microscopic scales are likely not too different 
than the system size.  This raises several questions. 
Is it theoretically consistent to apply hydrodynamics to systems whose size is of order 
microscopic scales?  Is it consistent to neglect nonhydrodynamic degrees of 
freedom when gradients are large?  What
is the size of the smallest drop of liquid?  

Since microscopic length and time scales typically decrease in the limit of strong coupling, 
it is natural to expect the domain of utility of hydrodynamics to be maximized at strong coupling.   
However, strongly coupled dynamics in QCD are notoriously difficult to study.
Holographic duality \cite{Maldacena:1997re} maps the dynamics 
of certain strongly coupled non-Abelian gauge theories onto 
the dynamics of classical gravity in one higher dimension.  The process of quark-gluon plasma
formation is mapped onto gravitational collapse and black hole formation with the ring down of the black hole
encoding the relaxation of the plasma to a hydrodynamic description.  Since 
gravitational collapse can be studied numerically, holography provides 
a unique arena to study all stages of evolution --- from far-from-equilibrium dynamics to hydrodynamics --- 
in a microscopically complete and controlled setting.   This 
has prompted much interest in using holographic theories as toy models of 
real quark-gluon plasma (for a review see \cite{CasalderreySolana:2011us}).
The simplest theory with a dual holographic description is $\mathcal N = 4$ 
supersymmetric Yang-Mills theory (SYM), which is dual to gravity in AdS$_5$.

A simple model of quark-gluon plasma production is the collision of 
gravitational shock waves
\cite{Grumiller:2008va,Albacete:2008vs,Albacete:2009ji,Chesler:2009cy,Chesleri,Casalderrey-Solana:2013aba,vanderSchee:2013pia,Bantilan:2012vu,Bantilan:2014sra,Chesler:2015wra},
which can result in the formation of a black hole.
The profile of the gravitational waves encodes the 
profile of colliding shock waves in the dual field theory.
In this Letter we report on shock wave collisions in SYM, which superficially at least,
resemble proton-nucleus collisions, with the ``nucleus"
represented as a planar shock wave and the ``proton" represented by  
shock wave localized in the transverse directions.  
We use quotes here to emphasize to the reader that the objects we are colliding 
are not bound states in QCD, but rather caricatures in SYM.  
The precollision geometry contains a trapped surface and the collision 
results in the formation of a black brane.
We numerically solve the full 5D Einstein equations for the 
geometry after the collision and report on the evolution of the 
SYM stress tensor $ T^{\mu \nu}$ and test 
the validity of hydrodynamics.  

In strongly coupled SYM the hydrodynamic gradient expansion of $T^{\mu \nu}$ is 
an expansion in powers of $1/(\ell T_{\rm eff})$ with $\ell$ the characteristic scale over which
$T^{\mu \nu}$ varies and $T_{\rm eff}$ the effective 
temperature \cite{Bhattacharyya:2008jc,Baier:2007ix}.  
Hence in strongly coupled SYM the microscopic scale $1/T_{\rm eff}$ plays the role of the mean free path.
We focus on the low energy limit, where $T_{\rm eff}$ is small and $1/T_{\rm eff}$ is large, and 
demonstrate that collisions can result in the formation of droplets of liquid with sizes as small as 
$R \sim 1/T_{\rm eff}$.
Our results bolster the notion that  debris produced in proton-nucleus collisions 
may be modeled using hydrodynamics.

\noindent \section{{Gravitational formulation}}
We construct initial data for Einstein's equations by superimposing the metric
of gravitational  shock waves moving in the $\pm z$ directions at the speed of light.
In Fefferman-Graham coordinates the metric of a single shock moving in the $\pm z$ direction is
\begin{align}
    ds^2 &=  r^2 \big[
	{-} dt^2 + d\bm x^2 + {\textstyle \frac{dr^2}{r^4}}
	+ h_\pm(\bm x_\perp, z_\mp,r) \, dz_\mp^2
    \big] \,,
\label{eq:FG}
\end{align}
where $\bm x \equiv \{x,y,z\}$, $\bm x_\perp \equiv \{x,y\}$, $z_\mp \equiv z \mp t$, and
\begin{align}
    h_\pm(\bm x_\perp,z_\mp,r) &\equiv
     \int \frac{d^2 k}{(2\pi)^2} \>
    e^{i {\bm k} \cdot \bm x_\perp} \,
    \widetilde H_\pm({\bm k},z_\mp) \,
    \frac{8I_2(k/r)}{k^2 r^2} \,.
\label{eq:h}
\end{align}
%The AdS curvature scale has been set to unity.
The boundary of the  spacetime lies at $r = \infty$.
The metric (\ref{eq:FG}) is an exact solution to Einstein's 
equations for any choice of $\widetilde H_\pm$ \cite{Gubser:2008pc,Grumiller:2008va}.
This geometry represents a state in the dual SYM theory with stress-energy tensor
\footnote
    {%
    In Eq.~(\ref{eq:singleshock}), and thereafter,
    $T^{\mu\nu}$ is really the expectation value of the SYM stress-energy tensor
    divided by $\Nc^2/(2 \pi^2)$, with $\Nc$ the gauge group rank
    \cite{deHaro:2000xn}.  In computing the ``proton" energy below, we use $N_{\rm c} = 3$.
    }

\begin{equation}
\label{eq:singleshock}
    T^{00} = T^{zz} = \pm T^{0z} = H_\pm(\bm x_\perp,z_\mp),
\end{equation}
(and all other components vanishing),
where $H_\pm$ is the  transverse Fourier transform of $\widetilde H_\pm$.
We choose the waves moving in the $+z$ and $-z$ directions to represent the 
``proton" and ``nucleus" respectively. 

A simple choice of shock profiles  is
\begin{equation}
H_\pm(\bm x_\perp,z_\mp) = \mu_{\pm}(\bm x_\perp)^3 \delta_w(z_\mp)
\end{equation}
where $\delta_w(z_\mp)$ is a smeared delta function with width $w$.  On the boundary the longitudinally 
integrated energy density per unit area is  $\mu_\pm(\bm x_\perp)^3$.
Note that under a boost in the $z$ direction (and in the $w\to0$ limit) 
only the normalization of $\mu_\pm$ 
change.  We may therefore work in the frame in which 
${\rm max}(\mu_+) = {\rm max}(\mu_-)$.  
% (with $\mu_\pm^3$
%the longitudinally integrated energy density of the ``proton" and ``nucleus").
For numerical convenience we chose
 $\mu_+(\bm x_\perp)^3 =  e^{-\frac 12 \bm x_\perp^2/\sigma^2}$ with $\sigma = 3$, and $\mu_-(\bm x_\perp)^3 = 1$.
Hence, the ``proton" stress is localized about $z=t, \ \bm x_\perp = 0$,
and the ``nucleus" stress is localized about $z=-t$ and is translationally invariant 
in the transverse plane.
Our choice of energy scale $\mu_\pm(\bm x_\perp {=} 0) =1$ fixes units in the results presented below.
For the smeared longitudinal $\delta$ function we use a Gaussian 
$\delta_w(z)=\frac{1}{\sqrt{2 \pi w^2}} e^{-\frac 12 z^2/w^2 }$ with width $w = 0.375$.
Note that evolution inside the future light cone of planar shock collisions with width $w = 0.375$ 
well approximates that of the $\delta$ function limit \cite{DeltaLimit}.  
%Fixing units by 
%setting $\sigma$ equal to the proton radius, $\sigma = 0.8768$ fm, in the CM frame our ``proton" 
%has energy $17.6$ GeV.

For early times, $t \ll -w$, the  profiles 
$H_\pm$ have negligible overlap and
the precollision geometry can be constructed from (\ref{eq:FG})
by replacing the last term with the sum of corresponding terms from
left and right moving shocks.
The resulting metric satisfies Einstein's equations, at early times,
up to exponentially small errors.

\begin{figure*}[ht]
\suck[scale = 0.4]{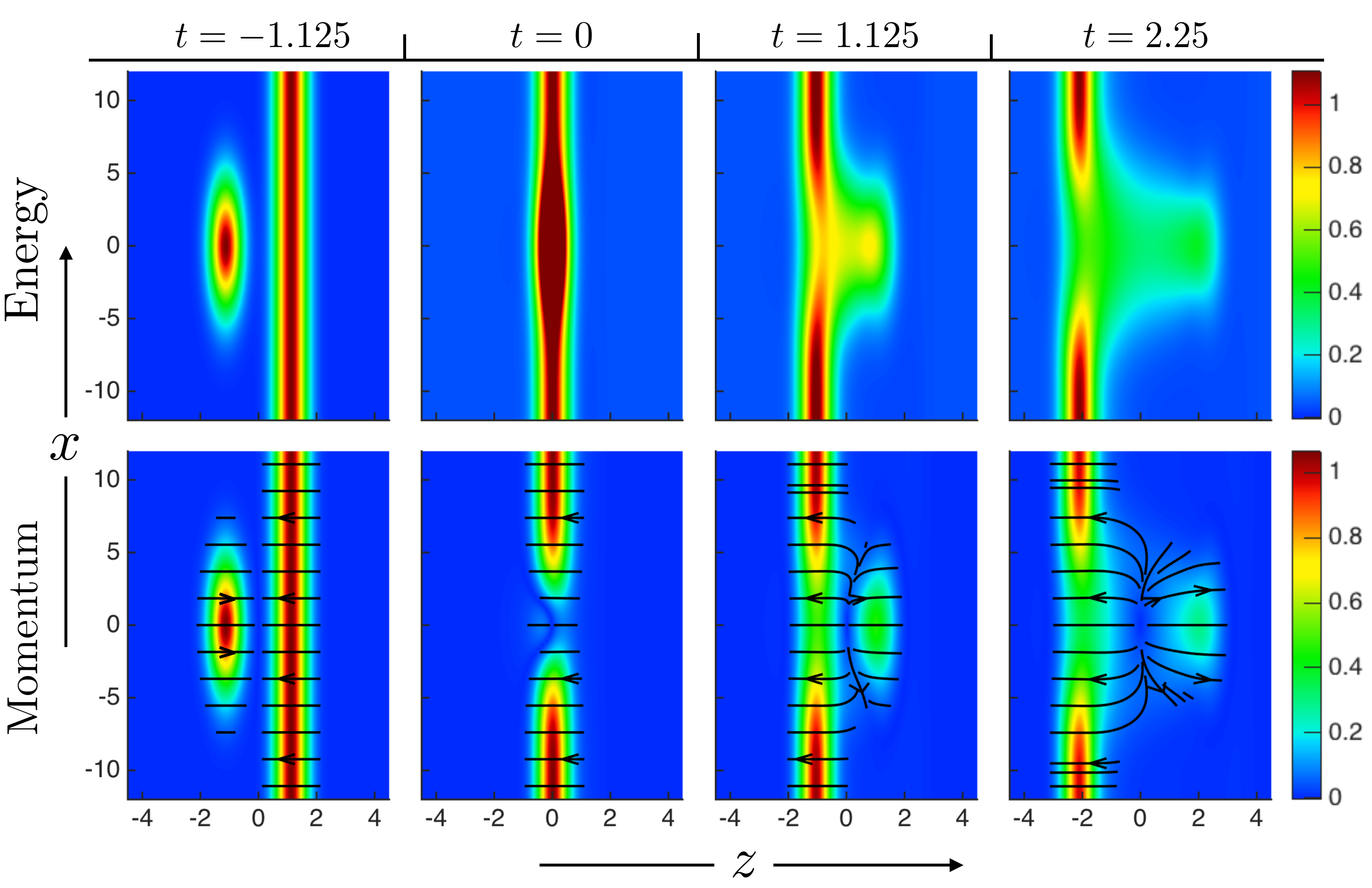}
%\vskip -0.05in
\caption{The  energy density $T^{00}$ (top) and momentum density $|T^{0i}|$ (bottom),
at four different times, in the plane $y = 0$.  
Streamlines in the lower plots
denote the direction of $T^{0i}$.  
The stress is rotationally invariant about the $z$ axis.
At time $t = -1.125$ the ``nucleus" is at $z = 1.125$ and the ``proton" is at 
$z = -1.125$.
These two distributions of energy move 
towards each other at the speed of light and collide at $t = z = 0$. 
Note that the color scaling for the energy at $t = 0$ is off scale.
Far away from $x = 0$, where there is little overlap between
the ``proton" and the ``nucleus," the collision has little effect on the future evolution of the ``nucleus."  
However, near $x = 0$ the ``proton" punches a hole in the ``nucleus." 
The resulting debris has transverse size of order that of the ``proton,"  lies 
inside the light cone and  expands  in the transverse and 
longitudinal directions.
\label{fig:snapshots}
} 
\end{figure*}

To evolve the precollision geometry forward in time we 
use the characteristic formulation of gravitational dynamics in
asymptotically AdS spacetimes discussed in detail in \cite{Chesler:2013lia}.
Our metric ansatz reads
\begin{equation}
    ds^2 = r^2 \, g_{\mu \nu}(x,r) \, dx^\mu dx^\nu + 2 \, dr \, dt \,,
\label{eq:ansatz}
\end{equation}
with Greek indices denoting spacetime boundary coordinates, $x^\mu = (t,x,y,z)$.
Near the boundary,
$g_{\mu \nu} = \eta_{\mu \nu}  + g_{\mu \nu}^{(4)}/r^4 + O(1/r^5)$.
The subleading coefficients $g^{(4)}_{\mu\nu}$ determine 
the SYM stress tensor,
\begin{equation}
%\label{eq:holostress}
T^{\mu \nu} = g_{\mu \nu}^{(4)}  + \tfrac{1}{4} \, \eta_{\mu \nu}\, g_{00}^{(4)}.
\end{equation}

To generate initial data for our characteristic evolution,
we numerically transform the precollision metric in
Fefferman-Graham coordinates 
to the metric ansatz (\ref{eq:ansatz})
\footnote
  {
   We also modify the initial data by adding a small uniform background energy density,
   equal to 4\% of the peak energy density of the incoming shocks. 
   This allows use of a coarser grid, reducing memory requirements.
  }.
We periodically compactify spatial directions with transverse size $L_x = L_y  = 24$
and longitudinal length $L_z = 9$.  We begin time evolution at $t = -1.5$
and evolve to $t = 2.5$.
Time evolution is performed using a spectral grid of size $N_x = N_y = 39$, $N_z = 155$ and $N_r = 48$.
%$39 \times 39 \times 145 \times 40$.

\noindent \section{{Results}} 
In Fig.~\ref{fig:snapshots} we plot the energy density $T^{00}$ (top)
and momentum density $T^{0i}$ (bottom)
in the plane $y = 0$ at several values of time.   (Note that the plots are rotationally invariant about the $z$ axis.)
The color scaling in the lower plots denotes $| T^{0i}|$ and the 
flow lines indicate the direction of $T^{0i}$.
At time $t = -1.125$ the system consists of a planar sheet of energy (the ``nucleus") localized at $z = 1.125$
and a localized lump of energy (the ``proton") centered at $z = -1.125$, $x = 0$, with transverse width $\sigma = 3$.  
These two distributions of energy move 
towards each other at the speed of light and collide at $t = z = 0$.   
Far away from $x = 0$, where there is little overlap between
the ``proton" and the ``nucleus," the collision has little effect on the future evolution of the ``nucleus."  
Indeed, at $x = \pm12$ the ``nucleus" simply continues to propagate at the speed of light with little change 
in the profile of the energy or momentum densities.  However, at $|x| \sim \sigma$
both the ``proton" and ``nucleus"  are dramatically altered 
by the collision event; the collision results in the ``proton" punching a hole of transverse size $\sim \sigma$
in the ``nucleus."  The resulting produced debris lies 
inside the light cone and subsequently expands both in the transverse and 
longitudinal directions.
As we argue below, it is in the region $|x| \sim \sigma$ --- set by the transverse size of the ``proton" --- that the 
system begins to behave hydrodynamically after the collision.

\begin{figure}
\suck[scale = 0.26]{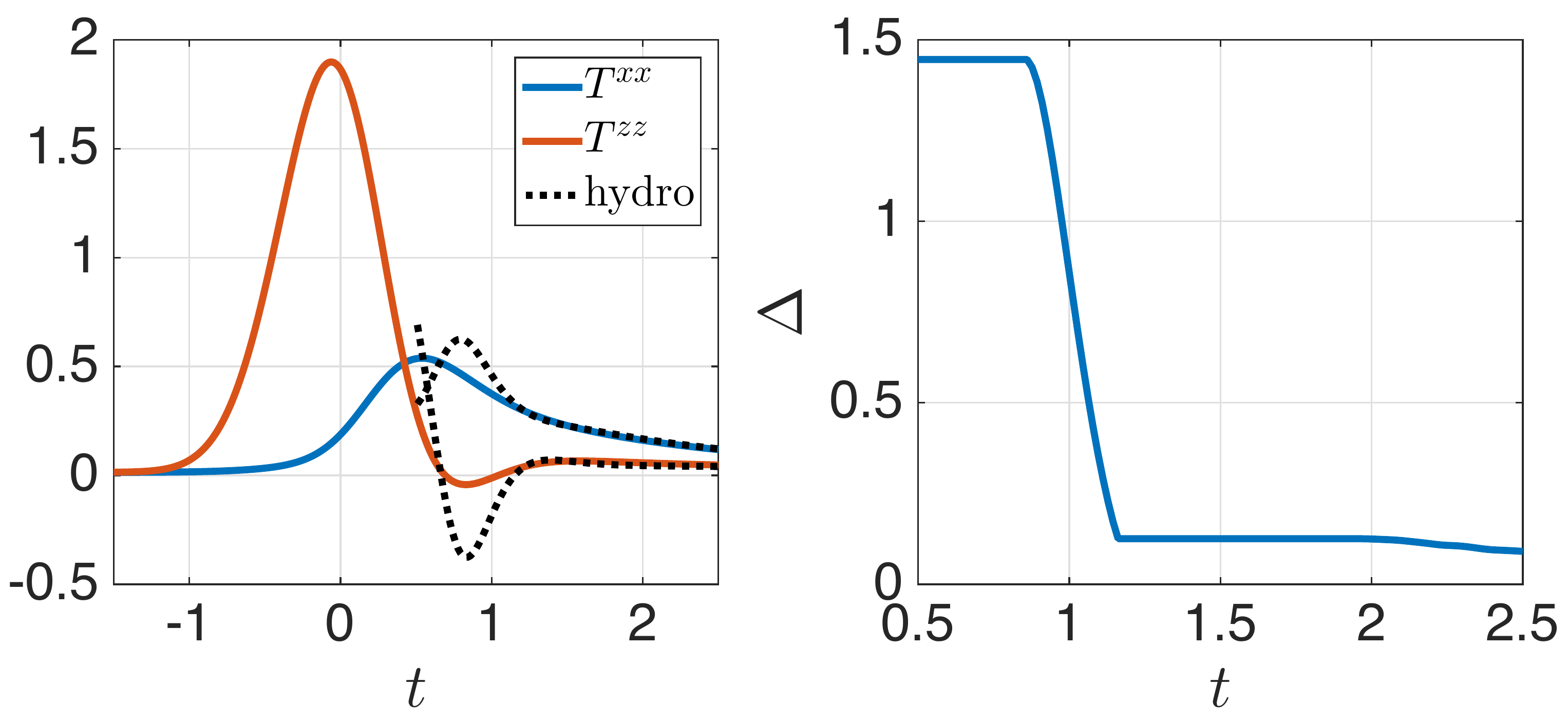}
%\vskip -0.05in
\caption{Left: Stress tensor components $ T^{xx}$ and $ T^{zz}$ 
at $\bm x = 0$ as a function of time together with 
their hydrodynamic approximation.
Around $t = 0$ the system is highly anisotropic
and far from equilibrium.  Nevertheless, at this point in space, the system begins to 
evolve hydrodynamically at $t \approx 1.2$.  Right: the hydrodynamic residual $\Delta$
at $\bm x = 0$ as a function of time.  After $t = 1.2$, $\Delta < 0.13$.
\label{fig:pressures}
} 
\end{figure}

At time and length scales $\gg$ than the microscopic scale $1/T_{\rm eff}$,
nonhydrodynamic degrees of freedom must relax, the evolution of 
$T^{\mu \nu}$ must be governed by hydrodynamics, and the dual black hole geometry must be governed by fluid/gravity duality \cite{Bhattacharyya:2008jc,Baier:2007ix}.
The salient hydrodynamic variables are the  fluid velocity $u^\mu$ and the proper energy 
$\epsilon$, which near equilibrium are related to $T_{\rm eff}$ by
\begin{equation}
\label{eq:Teff}
T_{\rm eff} = \left ( \textstyle \frac{4 \epsilon}{3 \pi^4} \right)^{1/4}.
\end{equation}
%With  $\epsilon$ and $u^\mu$ known, 
%the hydrodynamic approximation to the 
%stress tensor $T_{\rm hydro}^{\mu \nu}$ can be constructed using the constitutive relations
%of relativistic hydrodynamics.
%
To obtain $\epsilon$ and $u^\mu$, we extract the eigenvalues $p_{(\lambda)}$ and 
associated eigenvectors $e_{(\lambda)}^\mu$ of $T^{\mu \nu}$,
\begin{equation}
T^{\mu}_{\ \nu} \, e^\nu_{(\lambda)} = p_{(\lambda)} \, e^\mu_{(\lambda)} \,,
\label{eq:veldef}
\end{equation}
with no sum over $\lambda$ implied.
Near equilibrium $T^{\mu}_{\ \nu}$ has one timelike eigenvector, $e_{(0)}^\mu$,
and three spacelike eigenvectors, $e_{(i)}^\mu$.  The temporal eigenvalue 
is simply the proper energy, $\epsilon \equiv -p_{(0)}$,
with the local fluid velocity the associated eigenvector, $u^\mu \equiv e_{(0)}^\mu$.
We choose normalization $u_\mu u^\mu = -1$ with $u^0 > 0$.  The 
spatial eigenvalues $p_{(i)}$ represent 
local pressures in the fluid rest frame.  
With the complete spacetime dependence of the flow field $u^\mu$ and energy density $\epsilon$ determined from the exact stress, 
we then construct the hydrodynamic approximation to the stress tensor $T^{\mu \nu}_{\rm hydro}$
using the constitutive relations of $2^{\rm nd}$ order conformal hydrodynamics 
with transport coefficients determined by fluid/gravity duality \cite{Bhattacharyya:2008jc,Baier:2007ix}.

In Fig.~\ref{fig:pressures} we plot the stress tensor components $ T^{xx}$ and $T^{zz}$,
and their hydrodynamic approximations 
at $\bm x = 0$ as a function of time.
At this point $\bm u = O(10^{-2})$, the stress tensor is approximately diagonal,
and the local fluid pressures 
are well approximated by 
$T^{xx}$ and $T^{zz}$. 
The pressures increase dramatically during the collision, reflecting a 
system which is highly anisotropic and far from equilibrium.
However, after time $t = t_{\rm hydro} \sim 1.2$ there is a qualitative change in the behavior of $T^{xx}$ and $T^{zz}$;
thereafter the stress slowly varies in time and slowly isotropizes.  Likewise, after $t_{\rm hydro}$ 
the pressures are well described  the hydrodynamic constitutive relations.

\begin{figure*}[ht]
\suck[scale = 0.5]{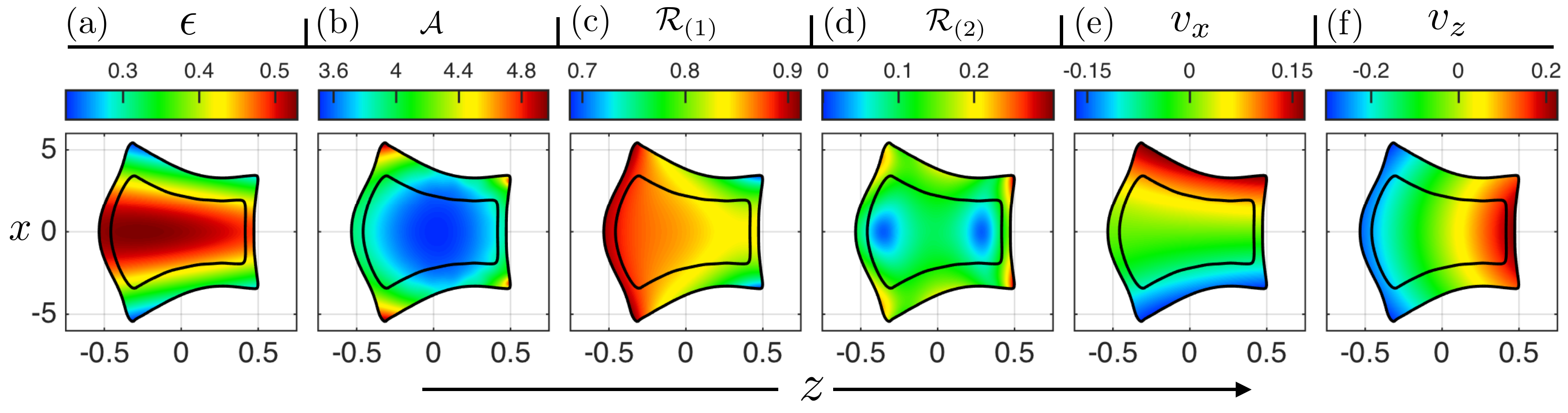}
%\vskip -0.05in
\caption{From left to right: 
the proper energy $\epsilon$, the anisotropy $\mathcal A$, 
the first and second order gradient measures $\mathcal R_{(1)}$ and $\mathcal R_{(2)}$,
and the transverse and longitudinal fluid velocities $v_x$ and $v_z$ respectively.
All plots are shown at time $t = 1.5$ and are restricted to the domain $\Delta \leq 0.2$ and are rotationally invariant about the $z$ axis.
The outer (inner) contours in all plots are $\Delta= 0.2\,(0.15)$,
with $\Delta< 0.2 \,(0.15)$ everywhere inside.  In the displayed region, which has transverse radius $R \sim 1/T_{\rm eff}$, hydrodynamics is a good description
of the evolution of the stress.  
%The observation that $\mathcal A \sim 4$ reflects the fact that gradients are large.
%Note that the transverse velocity is nearly as large as the longitudinal velocity.
%
\label{fig:DeltaPlots}
} 
\end{figure*}

To  elaborate on the domain of utility of hydrodynamics we
define the dimensionless residual measure 
%\footnote
%  {
%  Note $\Delta T^{\mu \nu} \Delta T_{\mu \nu}$ is in general
%  non-monotonic in time due to oscillations of nonhydrodynamic modes such as 
%  quasinormal modes.  Taking the max in (\ref{eq:deltadef}) 
%  ameliorates the effect of the oscillations and aids in avoiding 
%  the false identification of hydrodynamic evolution when $\Delta T^{\mu \nu} \Delta T_{\mu \nu}$
%  is temporarily small.
%  }
\begin{equation}
\label{eq:deltadef}
\Delta(t,\bm x) \equiv \underset{t' \ge t}{\max} \left (  \textstyle \frac{1}{\bar p(t',\bm x)} \sqrt{\Delta T_{\mu \nu}(t',\bm x) \Delta T^{\mu \nu}(t',\bm x)} \right ),
\end{equation}
with $\Delta T^{\mu \nu} \equiv T^{\mu \nu} - T^{\mu \nu}_{\rm hydro}$ and 
$\bar p \equiv \frac{1}{3} \sum_i p_{(i)} = \frac{\epsilon}{3}$ the average 
pressure.  In the local fluid rest frame $\Delta T^{\mu \nu} \Delta T_{\mu \nu} = \Delta T^{ij} \Delta T_{ij}$.
%\footnote
 % {
  Note that $\Delta T^{\mu \nu} \Delta T_{\mu \nu}$ is in general
  nonmonotonic in time due to nonhydrodynamic modes, such as 
  quasinormal modes, oscillating in time while decaying. 
  Indeed, $T^{xx}$ and $T^{zz}$ in Fig.~\ref{fig:pressures} agree with
  the hydrodynamic constitutive relations at $t \approx 0.65$, but 
  subsequently disagree until 
 $t \gtrsim 1.2$.
  Taking the max in (\ref{eq:deltadef}) 
  ameliorates the effect of the oscillations and aids in avoiding 
  the false identification of hydrodynamic evolution if $\Delta T^{\mu \nu} \Delta T_{\mu \nu}$
  is temporarily small.
%  }
If  $\Delta \ll 1$ at  $t,\bm x$,
then $T^{\mu \nu}$ is well described by the hydrodynamic constitutive relations at $\bm x$ at time $t$ and
all later times.
Also included in Fig.~\ref{fig:pressures} is a plot of $\Delta$ at $\bm x = 0$
as a function of time.  During the collision $\Delta \sim 1$ and $T^{\mu \nu}$ is not  described by hydrodynamics.
However, after $t \sim 0.8$,
$\Delta$ rapidly decays and after $t =1.2$, $\Delta < 0.13$.
In what follows we use $\Delta \lesssim 0.2$ as a litmus test
for whether the evolution of $T^{\mu \nu}$ is consistent with the hydrodynamic constitutive relations.

In the remaining plots Figs.~\ref{fig:DeltaPlots}(a)-\ref{fig:DeltaPlots}(f)
we restrict our attention to time $t = 1.5$ and to the region in the $x{-}z$ plane in where
$\Delta\leq 0.2$.  We explain the coloring of the different plots below.
The stress in the omitted regions has $\Delta > 0.2$ and is not well approximated by the hydrodynamic constitutive relations.
All plots in Fig.~\ref{fig:DeltaPlots} include the same
$\Delta= $ const. contours. The outer (inner) contour is $\Delta= 0.2\,(0.15)$,
with $\Delta< 0.2 \,(0.15)$ everywhere inside.   
The transverse radius of these contours is $R\sim 4$, which is approximately that of the initial
``proton" seen in Fig.~\ref{fig:snapshots}.  
We therefore conclude that the collision 
results in the formation of a droplet of liquid with transverse size of order that 
of the ``proton."   
The color scaling in Fig.~\ref{fig:DeltaPlots}(a) denotes the 
proper energy $\epsilon$.
In the plotted domain the average 
$\epsilon$ is $\sim 0.4$ and, via (\ref{eq:Teff}), the average $T_{\rm eff}$ is  $\sim 0.25$. 
We therefore obtain the dimensionless measure of the transverse size of the drop of liquid,
\begin{equation}
\label{eq:dimlesssize}
R T_{\rm eff} \sim 1.
\end{equation}
Evidently, hydrodynamics works even when the 
system size is of order the microscopic scale $1/T_{\rm eff}$.
Additionally, note $t_{\rm hydro} T_{\rm eff} \sim 0.3$.
Similar hydrodynamization times were observed in the 
1{+}1 dimensional flows of 
\cite{Chesler:2009cy,Casalderrey-Solana:2013aba,Heller:2013fn}
with $t_{\rm hydro} T_{\rm eff} \sim 0.3$ appearing as a lower bound.

When $R \sim 1/T_{\rm eff}$ it is natural to expect
gradients to be large.  
In order to quantify the size of gradients we define the anisotropy function 
\begin{equation}
\mathcal A \equiv {\max (p_{(i)})}/ {\min (p_{(i)})}.
\end{equation}
In ideal hydrodynamics, where the stress is isotropic in the local rest frame, 
$\mathcal A = 1$.  Therefore, deviations of $\mathcal A$ from $1$ in regions where $\Delta \ll 1$
must be due to gradient corrections to the hydrodynamic constitutive relations.
In Fig.~\ref{fig:DeltaPlots}(b) we plot 
$\mathcal A$.
In the displayed domain $\mathcal A \sim 4$.  Evidently, gradients are  large and ideal hydrodynamics is 
not a good approximation.

However, the observed anisotropy is almost entirely due to first order viscous effects
alone.  To see this, let $\Pi^{\mu \nu}_{(n)}$ be the $n^{\rm th}$ order contribution to the viscous stress
and define the dimensionless gradient measures 
\begin{equation}
\mathcal R_{(n)} \equiv \textstyle \frac{1}{\bar p} \sqrt{\Pi_{(n) \mu \nu} \Pi^{\mu \nu}_{(n)}}.
\end{equation}
In the local rest frame  $\Pi_{(n) \mu \nu} \Pi^{\mu \nu}_{(n)} = \Pi_{(n) ij} \Pi^{ij}_{(n)}$.
$\mathcal R_{(n)}$ measures the size of the $n^{\rm th}$ order gradients corrections
relative to ideal hydrodynamics.  In Figs.~\ref{fig:DeltaPlots}(c)-\ref{fig:DeltaPlots}(d) we plot $\mathcal R_{(1)}$
and  $\mathcal R_{(2)}$.  Inside the $\Delta = 0.15$ contour 
$\mathcal R_{(1)} \sim 0.85$ and $\mathcal R_{(2)} \sim 0.1$.  
Hence, first order gradient corrections
are as large as the ideal stress with the second order correction an order of magnitude smaller.  
Similar observations were made in \cite{Chesler:2009cy,Heller:2013fn} 
within the context boost invariant flow.

\noindent \section{{Discussion}}%
Given that $1/T_{\rm eff}$ is the salient microscopic scale in strongly coupled plasma ---  akin to a mean free path at weak coupling --- 
it is remarkable that hydrodynamics can describe the evolution systems as small as $R \sim 1/T_{\rm eff}$.
%Simply put, even when $R \sim 1/T_{\rm eff}$ and gradients are large, nonhydrodynamic modes
%decay and the subsequent hydrodynamic evolution is self-consistent.
Simply put, even when $R \sim 1/T_{\rm eff}$ and gradients are large, nonhydrodynamic modes
decay and the hydrodynamic gradient expansion is well behaved.  

It should be noted that
upon fixing units by setting $\sigma$ equal to the proton radius, $\sigma \sim 1$ fm, in the c.m. frame our ``proton" 
has energy $\sim 20$ GeV and the effective temperature of the produced plasma is $T_{\rm eff} \sim 200$ MeV.  Given that 
proton-nucleus collisions at RHIC and the LHC have energies and temperatures greater than this, it is natural to expect 
$R T_{\rm eff}$ to be larger in RHIC and LHC collisions than the simulated collision presented here.  Given that hydrodynamics already works well
when $R T_{\rm eff} \sim 1$, our observed extreme applicability of hydrodynamics, both in terms of the system size and 
the size of gradients, bolsters the notion the debris produced in proton-nucleus
collisions can be modeled with hydrodynamics.  

However, hydrodynamic simulations of 
tiny short lived systems with large gradients will likely be sensitive to the initial viscous stress, 
which Figs.~\ref{fig:DeltaPlots}(b)-\ref{fig:DeltaPlots}(c) demonstrate is large, 
and to the initial transverse fluid velocity. 
Indeed, as shown in Fig.~\ref{fig:DeltaPlots}(e)-\ref{fig:DeltaPlots}(f), the transverse and longitudinal components of the 
fluid 3-velocity $\bm v \equiv \bm u/u^0$ are similar in magnitude.
Simply put, large gradients result in large viscous stress and drive the rapid development 
of transverse flow. These effects are likely key ingredients in initial hydrodynamic data required for precision 
modeling of proton-nucleus collisions.

It would be interesting to push our analysis further: how big are the smallest drops of liquid?
Clearly $R T_{\rm eff}$ cannot be made arbitrarily small since the 
hydrodynamic gradient expansion is badly behaved when $R T_{\rm eff} \ll 1$.
%When the ``proton" energy $E$ decreases, does the dimensionless 
%size $R T_{\rm eff}$ of the produced droplet of liquid also decrease?  
%Is there a lower bound
%on $R T_{\rm eff}$ or does the hydrodynamic description simply 
%become progressively worse with the hydrodynamization time $t_{\rm hydro}$ increasing as $E$ decreases?
Additionally, in the gravitational description %, critical gravitational collapse \cite{Choptuik:1992jv} dictates 
there should exist a critical energy $E_{\rm c}$, below which no black hole is formed, and which for
$E = E_{\rm c} + 0^+$ critical gravitational collapse occurs \cite{Choptuik:1992jv}.
The absence of a black hole when $E<E_{\rm c}$ means that in the dual field theory 
the collisional debris will not evolve hydrodynamically at any future time.  It would be interesting 
to study dynamics near $E_c$ and the signatures of hydrodynamics turning off.

%\begin{acknowledgments}
{\textit{Acknowledgments.}---}I am grateful to Janet Johnson, Krishna Rajagopal and Subir Sachdev for assistance 
with computer resources required to complete this project.  
This work is supported
by the Fundamental Laws Initiative of the Center for the
Fundamental Laws of Nature at Harvard University.
%\end{acknowledgments}

\bibliographystyle{utphys}
\bibliography{refs}%
\end{document}